\documentclass[final,3p,times,numbers,sort&compress]{elsarticle}

\usepackage{amsmath}
\usepackage{amssymb}
\usepackage{amsthm}
\usepackage{mathtools}
\usepackage[utf8]{inputenc}
\usepackage{bm}
\usepackage{color}
\usepackage{xcolor}
\usepackage{soul}

\journal{Physica A}

\usepackage[breaklinks=true,colorlinks=true,linkcolor=blue,urlcolor=blue,citecolor=blue]{hyperref}

\newcommand{\defeq}{\mathrel{\mathop:}=}

\newcommand{\pS}[1]{P(S|#1)}
\newcommand{\pL}[1]{P(L|#1)}

\begin{document}

\begin{frontmatter}

\title{Configurational density of states and melting of simple solids}

\address[cchen]{Research Center on the Intersection in Plasma Physics, Matter and Complexity, P$^2$MC,\\ Comisión Chilena de Energía Nuclear, Casilla 188-D, Santiago, Chile}
\address[unab]{Departamento de F\'isica, Facultad de Ciencias Exactas, Universidad Andres Bello. Sazi\'e 2212, piso 7, Santiago, 8370136, Chile.}
\author[cchen,unab]{Sergio Davis\corref{cor1}}
\author[unab]{Claudia Loyola}
\author[unab]{Joaquín Peralta}
\cortext[cor1]{Corresponding author}

\begin{abstract}
We analyze the behavior of the microcanonical and canonical caloric curves for a piecewise model of the configurational density of states of simple solids, in the context 
of melting from the superheated state, as realized numerically in the Z-method via atomistic molecular dynamics. A first-order phase transition with metastable regions is 
reproduced by the model, being therefore useful to describe aspects of the melting transition. Within this model, transcendental equations connecting the superheating limit, 
the melting point, and the specific heat of each phase are presented and numerically solved. Our results suggest that the essential elements of the microcanonical Z curves 
can be extracted from simple modeling of the configurational density of states.
\end{abstract}


\end{frontmatter}

\section{Introduction}

One of the widely used approaches to determine the melting point of materials via atomistic computer simulation is the so-called Z-method~\cite{Belonoshko2006,Belonoshko2007,Bouchet2009,
Li2011,Belonoshko2012,Stutzmann2015,GonzalezCataldo2016,Anzellini2019,Errandonea2020,Mausbach2020,Baty2021}, which is based on the empirical observation 
that the superheated solid at the limit of superheating temperature $T_{LS}$ has the same internal energy as the liquid at the melting temperature $T_m$. This observation has 
potential implications for understanding the atomistic melting mechanism from the superheated state~\cite{Davis2011,OlguinArias2019} but has been left mostly unexplored, disconnected 
from thermodynamical models of solids.

In the Z-method, the isochoric curve $T(E)$ is computed from simulations at different total energies $E_1, E_2, \ldots$ and the minimum temperature of the liquid branch of this 
isochoric curve is identified with the melting temperature $T_m$. This key assumption still lacks a proper explanation in terms of microcanonical thermodynamics of finite systems.

In this work, we study the properties of a recently proposed model~\cite{Montecinos2021} for the configurational density of states (CDOS) of systems with piecewise constant
heat capacity by calculating its canonical and microcanonical caloric curves in terms of special functions. The model presents a first-order phase transition with
metastable regions where the microcanonical curve $T(E)$ shows a so-called \emph{van der Waals loop}~\cite{Umirzakov1999}. The inflection points found in this loop can 
be associated to $T_{LS}$ and $T_m$ of the Z-method description of the superheated solid.

This paper is organized as follows. In Section \ref{sec:cdos} we briefly review the definition of the CDOS and the formalism used to compute thermodynamic properties from it. 
In Section \ref{sec:model} we revisit the model in Ref.~\cite{Montecinos2021} and provide some interpretation of its parameters. Sections ~\ref{sec:canon} and ~\ref{sec:microcanon} 
show the computation of the caloric curves of the solid model in the canonical and microcanonical ensemble, respectively, and we present some concluding remarks in 
Section ~\ref{sec:conclusions}.

\section{Configurational density of states and thermodynamics}
\label{sec:cdos}

\noindent
For a classical system with Hamiltonian
\begin{equation}
H(\bm{r}_1, \ldots, \bm{r}_N, \bm{p}_1, \ldots, \bm{p}_N) = \sum_{i=1}^N \frac{{\bm{p}_i}^2}{2 m_i} + \Phi(\bm{r}_1, \ldots, \bm{r}_N),
\end{equation}
we will define the configurational density of states (CDOS) as the multidimensional integral
\begin{equation}
\label{eq:cdos}
\mathcal{D}(\phi) \defeq \int d\bm{r}_1\ldots d\bm{r}_N\: \delta(\Phi(\bm{r}_1, \ldots, \bm{r}_N)-\phi),
\end{equation}
where $\Phi(\bm{r}_1, \ldots, \bm{r}_N)$ is the potential energy describing the interaction between particles. Using the definition of $\mathcal{D}(\phi)$
in (\ref{eq:cdos}) is possible to rewrite any configurational integral of the form
\begin{equation}
I = \int d\bm{r}_1\ldots d\bm{r}_N\:G(\Phi(\bm{r}_1, \ldots, \bm{r}_N))
\end{equation}
as a one-dimensional integral over $\phi$. In fact, taking $I$ and introducing a factor of 1 as an integral over a Dirac delta function, we have
\begin{equation}
\label{eq:cdos_trick}
\begin{split}
I & = \int d\bm{r}_1\ldots d\bm{r}_N\:G(\Phi(\bm{r}_1, \ldots, \bm{r}_N)) \\
  & = \int d\bm{r}_1\ldots d\bm{r}_N\left[\int_{-\infty}^\infty d\phi\delta(\phi-\Phi(\bm{r}_1,\ldots,\bm{r}_N))\right]\:G(\Phi(\bm{r}_1, \ldots, \bm{r}_N)) \\
  & = \int_{0}^\infty d\phi\left\{\int d\bm{r}_1\ldots d\bm{r}_N\delta(\phi-\Phi(\bm{r}_1,\ldots,\bm{r}_N))\right\}\:G(\phi) \\
  & = \int_{0}^\infty d\phi \mathcal{D}(\phi)G(\phi).
\end{split}
\end{equation}
where we have assumed that $\Phi$ has its global minimum at $\Phi = 0$. One such integral of particular importance is the canonical partition function 
$Z(\beta)$, defined as
\begin{equation}
Z(\beta) \defeq \int d\bm{\Gamma}\exp\big(-\beta H(\bm \Gamma)\big),
\end{equation}
where $\bm{\Gamma} = (\bm{r}_1,\ldots, \bm{r}_N, \bm{p}_1, \ldots, \bm{p}_N)$, and which can be computed, following \eqref{eq:cdos_trick}, as
\begin{equation}
\label{eq:zeta}
\begin{split}
Z(\beta) = & \int d\bm{p}_1\ldots d\bm{p}_N \exp\Big(-\beta\textstyle\sum_{i=1}^N \frac{\bm{p}^2_i}{2 m_i}\Big)
            \times \int d\bm{r}_1\ldots d\bm{r}_N\:\exp\Big(-\beta\Phi(\bm{r}_1, \ldots, \bm{r}_N)\Big) \\
    = &\: Z_0(N)\:\beta^{-3N/2}\int_{0}^{\infty} d\phi \mathcal{D}(\phi)\exp(-\beta \phi) \\
    = &\: Z_0(N)\:\beta^{-3N/2}Z_c(\beta),
\end{split}
\end{equation}
with $Z_0(N) \defeq \prod_{i=1}^N(\sqrt{2\pi m_i})^3$ a constant only dependent on $N$ and the masses of the particles, and where
\begin{equation}
Z_c(\beta) \defeq \int_0^\infty d\phi \mathcal{D}(\phi)\exp(-\beta \phi)
\end{equation}
is the configurational partition function. Similarly, the full density of states,
\begin{equation}
\Omega(E) \defeq \int d\bm{\Gamma}\delta(E-H(\bm \Gamma))
\end{equation}
can also be computed from the CDOS, by a convolution with the density of states $\Omega_K$ of the ideal gas~\cite{Kardar2007}, that is,
\begin{equation}
\label{eq:convol}
\Omega(E) = \int d\phi\mathcal{D}(\phi)\Omega_K(E-\phi),
\end{equation}
where
\begin{equation}
\label{eq:kdos}
\Omega_K(E) \defeq \int d\bm{p}_1\ldots d\bm{p}_N\delta\Big(\textstyle\sum_{i=1}^N \frac{\bm{p}^2_i}{2 m_i} - E\Big) = \Omega_0(N) \Theta(E)E^{\frac{3N}{2}-1}.
\end{equation}

\noindent
In order to obtain the result in (\ref{eq:convol}), we replace the integral over the momenta in $\Omega(E)$ by using
(\ref{eq:kdos}),
\begin{equation}
\begin{split}
\Omega(E) \defeq & \int d\bm{r}_1\ldots d\bm{r}_N\:\left[\int d\bm{p}_1\ldots d\bm{p}_N \:\delta\Big(\textstyle\sum_{i=1}^N \frac{\bm{p}^2_i}{2 m_i} + \Phi(\bm{r}_1, \ldots, \bm{r}_N)-E\Big)\right] \\
          = & \int d\bm{r}_1\ldots d\bm{r}_N\:\Omega_K(E-\Phi(\bm{r}_1, \ldots, \bm{r}_N))\\
          = & \int_{0}^\infty d\phi\mathcal{D}(\phi)\Omega_K(E-\phi),
\end{split}
\end{equation}
where in the last equality we have used (\ref{eq:cdos_trick}). Finally we have
\begin{equation}
\Omega(E) = \Omega_0(N)\int_0^E d\phi\mathcal{D}(\phi)(E-\phi)^{\frac{3N}{2}-1} = \Omega_0(N)\eta(E)
\end{equation}
where $\Omega_0(N) \defeq Z_0(N)/\Gamma\big(3N/2\big)$ is a function only of the size $N$ of the system, and
\begin{equation}
\eta(E) \defeq \int_0^E d\phi \mathcal{D}(\phi)\big(E-\phi\big)^{\frac{3N}{2}-1}.
\end{equation}

Now we have all the elements needed for the computation of the caloric curves and the transition energy (or temperature) for a given model of CDOS. In the canonical 
ensemble, the probability of observing a value $\phi$ of potential energy at inverse temperature $\beta$ is given by
\begin{equation}
\label{eq:phi_prob_canon}
P(\phi|\beta) = \frac{1}{Z_c(\beta)}\exp(-\beta\phi)\mathcal{D}(\phi)
\end{equation}
and, using this, we can determine the caloric curve (internal energy as a function of inverse temperature) as
\begin{equation}
\big<H\big>_\beta = -\frac{\partial}{\partial \beta}\ln Z(\beta) = \frac{3N}{2\beta} + \big<\phi\big>_\beta,
\end{equation}
where
\begin{equation}
\label{eq:phi_beta}
\big<\phi\big>_\beta = \frac{1}{Z_c(\beta)}\int_0^\infty d\phi \mathcal{D}(\phi)\exp(-\beta \phi)\phi = -\frac{\partial}{\partial \beta}\ln Z_c(\beta).
\end{equation}

On the other hand, in the microcanonical ensemble the probability of having potential energy $\phi$ at total energy $E$ is given by~\cite{Pearson1985, Ray1991, Carignano2002, Davis2011a}
\begin{equation}
\label{eq:phi_prob_micro}
P(\phi|E) = \frac{1}{\eta(E)}(E-\phi)^{\frac{3N}{2}-1}\mathcal{D}(\phi).
\end{equation}

\noindent
The microcanonical caloric curve (inverse temperature as a function of internal energy) is given by
\begin{equation}
\label{eq:beta}
\beta(E) \defeq \frac{\partial}{\partial E}\ln \Omega(E),
\end{equation}
which can also be rewritten as an expectation as follows. Replacing $\Omega(E)$ in terms of $\eta(E)$ in (\ref{eq:beta}) and using (\ref{eq:phi_prob_micro}), we 
can write
\begin{equation}
\begin{split}
\beta(E) & = \frac{1}{\eta(E)}\frac{\partial}{\partial E}\int_0^E d\phi\mathcal{D}(\phi)(E-\phi)^{\frac{3N}{2}-1} \\
         & = \frac{1}{\eta(E)}\int_0^E d\phi \mathcal{D}(\phi)\big(E-\phi)^{\frac{3N}{2}-1}\left[\frac{3N-2}{2(E-\phi)}\right] = \big<\hat{\beta}_K\big>_E.
\end{split}
\end{equation}
where $\hat{\beta}_K$ is the kinetic inverse temperature estimator
\begin{equation}
\hat{\beta}_K(\phi; E) \defeq \frac{3N-2}{2(E-\phi)}.
\end{equation}

\section{Model for the CDOS}
\label{sec:model}

In the following sections we will use, for the configurational density of states, the model presented in Ref.~\cite{Montecinos2021} which is defined piecewise, as
\begin{equation}
\mathcal{D}(\phi) = \begin{cases}
d_S(\phi-\phi_S)^{\alpha_S} \qquad\text{for}\;\phi < \phi_c, \\[15pt]
d_L(\phi-\phi_L)^{\alpha_L} \qquad\text{for}\;\phi \geq \phi_c.
\end{cases}
\end{equation}

This model represents two segments, one for the solid phase which, by definition, will have potential energies $\phi < \phi_c$, and one for the liquid phase where 
$\phi \geq \phi_c$. That is, our main assumption is that the potential energy landscape is effectively divided into solid and liquid states by a surface 
\[\Phi(\bm{r}_1, \ldots, \bm{r}_N) = \phi_c\] in configurational space. By imposing continuity of the CDOS at $\phi=\phi_c$, we must have
\begin{equation}
d_S(\phi_c-\phi_S)^{\alpha_S} = d_L(\phi_c-\phi_L)^{\alpha_L}.
\end{equation}

Here $\phi_S$ represents the potential energy minimum of the ideal solid, that can be set to zero without loss of generality provided that all energies are
measured with respect to this value. We can also set $d_S$=1 and express the potential energy in units of $\phi_c$, and then we have
\begin{equation}
\label{eq:model_CDOS}
\mathcal{D}(\phi) = \begin{cases}
\phi^{\alpha_S} \qquad\text{for}\;\phi < 1, \\[15pt]
\displaystyle\Big(\frac{\phi-\gamma}{1-\gamma}\Big)^{\alpha_L} \qquad\text{for}\;\phi \geq 1,
\end{cases}
\end{equation}
where we have defined the dimensionless parameter \[\gamma \defeq \frac{\phi_L}{\phi_c}.\] In this way, the model for the CDOS has only three free parameters, namely 
$\alpha_S$, $\alpha_L$ and $\gamma$.

\section{The caloric curve in the canonical ensemble}
\label{sec:canon}

The configurational partition function $Z_c(\beta)$ associated to the model for the CDOS in \eqref{eq:model_CDOS} can be obtained by piecewise integration as
\begin{equation}
\begin{split}
Z_c(\beta) & = \int_0^\infty d\phi \mathcal{D}(\phi)\exp(-\beta \phi) \\
           & = \int_0^1 d\phi \phi^{\alpha_S}\exp(-\beta \phi) + \frac{1}{(1-\gamma)^{\alpha_L}}\int_1^\infty d\phi (\phi-\gamma)^{\alpha_L}\exp(-\beta \phi) \\
           & = \beta^{-(\alpha_S+1)}\int_0^{\beta} du u^{\alpha_S}\exp(-u) + \frac{\beta^{-(\alpha_L+1)}}{(1-\gamma)^{\alpha_L}}\int_{\beta}^{\infty} du (u-\beta\gamma)^{\alpha_L}\exp(-u)\\
           & = \beta^{-(\alpha_S+1)}\int_0^{\beta} du u^{\alpha_S}\exp(-u) + \frac{\beta^{-(\alpha_L+1)}}{(1-\gamma)^{\alpha_L}}\exp(-\beta\gamma)\int_{\beta(1-\gamma)}^{\infty} dw w^{\alpha_L}\exp(-w).
\end{split}
\end{equation}

\noindent
Finally we obtain
\begin{equation}
\label{eq:Zc}
Z_c(\beta) = \beta^{-(\alpha_S+1)}G_S(0 \rightarrow \beta) + \frac{\beta^{-(\alpha_L+1)}}{(1-\gamma)^{\alpha_L}}\exp(-\beta\gamma)G_L(\beta(1-\gamma) \rightarrow \infty)
\end{equation}
where we have defined, for convenience, the auxiliary functions
\begin{equation}
G_\nu(a \rightarrow b) \defeq \int_a^b dt \exp(-t)t^{\alpha_\nu} = \Gamma(\alpha_\nu+1; b) - \Gamma(\alpha_\nu+1; a)
\end{equation}
for $\nu = S, L$, where $\Gamma(k; x) \defeq \int_0^x dt\exp(-t)t^{k-1}$ is the lower incomplete Gamma function.

\subsection{Low and high-temperature limits}

\vspace{5pt}
By taking the limit $\beta \rightarrow \infty$ of (\ref{eq:Zc}) we see that the second term vanishes, and also $G_S(0 \rightarrow \infty) = \Gamma(\alpha_S+1)$ so
for low temperatures we can approximate
\begin{equation}
\label{eq:Zc_approx}
Z_c(\beta) \approx \beta^{-(\alpha_S+1)}\Gamma(\alpha_S+1).
\end{equation}

\noindent
By replacing (\ref{eq:Zc_approx}) in (\ref{eq:phi_beta}) we have
\begin{equation}
\big<\phi\big>_\beta = -\frac{\partial}{\partial \beta}\ln Z_c(\beta) \approx \frac{\alpha_S+1}{\beta}
\end{equation}
and then the solid branch of the canonical caloric curve is given by a straight line,
\begin{equation}
\label{eq:solid_branch}
E_S(T) \defeq \big<H\big>_{T,S} = \frac{3N}{2}k_B T + \big<\Phi\big>_T = \big(\textstyle{\frac{3N}{2}}+\alpha_S+1\big)k_B T
\end{equation}
as expected. Here we can verify that $E_S(T) \rightarrow 0$ as $T \rightarrow 0$, because we have fixed $\phi_S = 0$, and moreover, we learn that the value of the
specific heat for the solid phase is
\begin{equation}
C_S = \frac{d}{dT}E_S(T) = \big(\textstyle{\frac{3N}{2}}+\alpha_S+1\big)k_B.
\end{equation}

\noindent
On the other hand, for high temperatures (i.e. in the limit $\beta \rightarrow 0$) we can approximate
\begin{equation}
\label{eq:Zc_approx2}
Z_c(\beta) \approx \left[\frac{\beta^{-(\alpha_L+1)}\exp(-\beta\gamma)}{(1-\gamma)^{\alpha_L}}\right]\Gamma(\alpha_L+1),
\end{equation}
obtaining from (\ref{eq:phi_beta}) that
\begin{equation}
\big<\phi\big>_\beta = -\frac{\partial}{\partial \beta}\ln Z_c(\beta) \approx \frac{\alpha_L+1}{\beta}+\gamma
\end{equation}
hence the liquid branch is also a straight line, given by
\begin{equation}
E_L(T) \defeq \big<H\big>_{T,L} = \frac{3N}{2}k_B T + \big<\Phi\big>_T \approx \gamma + \big(\textstyle{\frac{3N}{2}}+\alpha_L+1\big)k_B T.
\end{equation}

This allows us to interpret $\gamma$ as the extrapolation of the liquid branch towards $T=0$, that is, $\phi_L$ in units of $\phi_c$ is the potential energy of a perfectly frozen
liquid at $T=0$,
\begin{equation}
E_{L}(T=0) = \gamma = \frac{\phi_L}{\phi_c}.
\end{equation}

\noindent
Moreover, we obtain that the specific heat of the liquid phase is
\begin{equation}
C_L = \frac{d}{dT}E_L(T) = \big(\textstyle{\frac{3N}{2}}+\alpha_L+1\big)k_B.
\end{equation}

In order for the energy to be extensive in both branches as $N \rightarrow \infty$, it must hold true that the parameters $\alpha_\nu$ are proportional to $N$, and it follows
that
\begin{equation}
\frac{C_\nu}{k_B} = \alpha_\nu + \frac{3N}{2}
\end{equation}
with $\nu = S, L$. Using these definitions we can write $E_S$ and $E_L$ in terms of $\beta$ more compactly, as
\begin{subequations}
\begin{align}
E_S(\beta) & = \frac{C_S}{\beta}, \\
E_L(\beta) & = \gamma + \frac{C_L}{\beta}.
\end{align}
\end{subequations}

\subsection{Melting temperature}

\begin{figure}
\begin{center}
\includegraphics[width=0.5\textwidth]{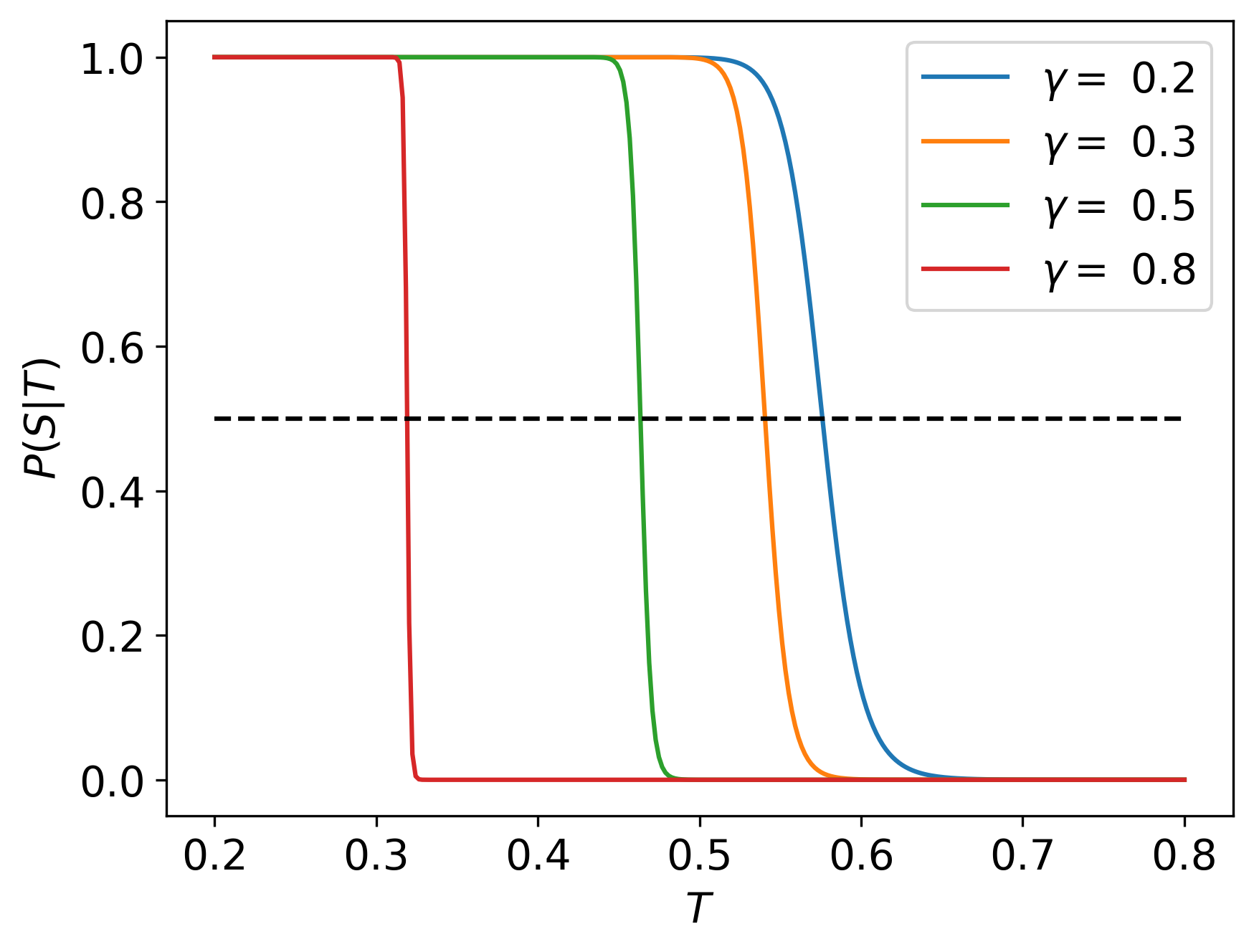}
\end{center}
\caption{Canonical probability of solid phase $P(S|T)$ as a function of temperature, for different values of $\gamma$.}
\label{fig:probcanon}
\end{figure}

On account of the assumption that all solid states have $\phi < \phi_c$ and for the liquid states $\phi > \phi_c$, we will define the probabilities $\pS{\beta}$ of being 
in the solid phase, and $\pL{\beta}$ of being in the liquid phase as
\begin{align}
\pS{\beta} \defeq P(\phi < 1|\beta) & = \frac{1}{Z_c(\beta)}\int_0^1 d\phi \exp(-\beta\phi)\phi^{\alpha_S} = \frac{\beta^{-(\alpha_S+1)} G_S(0\rightarrow \beta)}{Z_c(\beta)}, \\
\pL{\beta} \defeq P(\phi \geq 1|\beta) & = \frac{1}{Z_c(\beta)}\int_1^\infty d\phi \exp(-\beta \phi)\frac{(\phi-\gamma)^{\alpha_L}}{(1-\gamma)^{\alpha_L}}
                                       = \frac{\beta^{-(\alpha_L+1)}\exp(-\beta\gamma)}{Z_c(\beta)(1-\gamma)^{\alpha_L}}G_L(\beta(1-\gamma) \rightarrow \infty)
\end{align}
respectively, such that \[\pS{\beta} + \pL{\beta} = 1\] for all values of $\beta$. The probability of solid is shown as a function of $T$ in Fig.~\ref{fig:probcanon}. 
The melting temperature $T_m$ is such that both probabilities are equal~\cite{Davis2016}, that is,
\begin{equation}
\pS{\beta_m} = \pL{\beta_m} = \frac{1}{2},
\end{equation}
with $\beta_m = 1/(k_B T_m)$, which is then the solution of the transcendental equation
\begin{equation}
\label{eq:Tm_solution}
(\beta_m)^{\alpha_L-\alpha_S}G_S(0\rightarrow \beta_m) = \frac{\exp(-\beta_m\gamma)}{(1-\gamma)^{\alpha_L}}G_L(\beta_m(1-\gamma) \rightarrow \infty).
\end{equation}

\noindent
Using $\pS{\beta}$ and $\pL{\beta}$ we can write the canonical caloric curve for any $\beta$ as the sum of three contributions, namely
\begin{equation}
\label{eq:canon_energy_full}
\big<H\big>_\beta = \frac{3N}{2\beta} -\frac{\partial}{\partial \beta}\ln Z_c(\beta)
   = E_S(\beta)\pS{\beta} + E_L(\beta)\big(1-\pS{\beta}\big) + E_{\text{res}}(\beta),
\end{equation}
where the first and second terms account for the solid and liquid branches, respectively, and $E_\text{res}$ is equal to
\begin{equation}
E_{\text{res}}(\beta) \defeq -\frac{\gamma\beta^{-1}\exp(-\beta)}{Z_c(\beta)}.
\end{equation}

By comparing (\ref{eq:canon_energy_full}) with the low and high temperature limits, namely $E_S$ and $E_L$, and noting that $\pS{\beta} \rightarrow 1$ for low temperatures
and $\pS{\beta} \rightarrow 0$ for high temperatures, we see that $E_{\text{res}}$ must vanish on both limits, as can be verified by using (\ref{eq:Zc_approx}) and
(\ref{eq:Zc_approx2}). Therefore, this term is only relevant near the transition region. If we replace (\ref{eq:Tm_solution}) into the configurational partition
function in (\ref{eq:Zc}) we obtain
\begin{equation}
Z_c(\beta_m) = 2(\beta_m)^{-(\alpha_S+1)}G_S(0\rightarrow \beta_m)
\end{equation}
and we can determine the melting energy $E^*$ as
\begin{equation}
\label{eq:energy_star}
E^* = \frac{3N}{2\beta_m} + \frac{1}{2}\left[N\gamma+\frac{\alpha_S+\alpha_L+2}{\beta_m}\right] -\frac{\gamma\exp(-\beta_m)}{2G_S(0 \rightarrow \beta_m)}(\beta_m)^{\alpha_S}.
\end{equation}

\begin{figure}
\begin{center}
\includegraphics[width=0.5\textwidth]{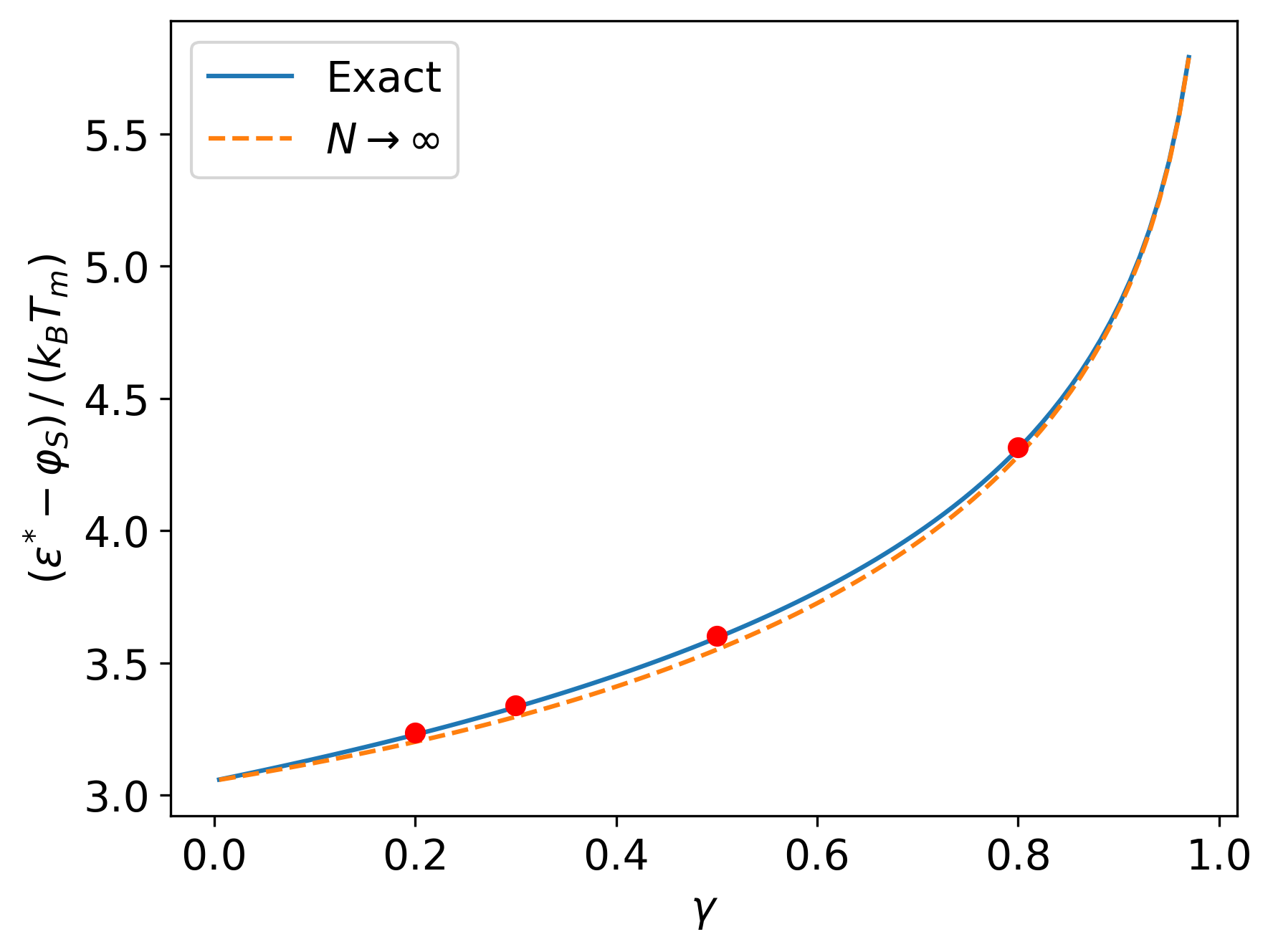}
\end{center}
\caption{Ratio between the melting energy and temperature, according to (\ref{eq:energy_star}) (solid blue line) and its approximation
(\ref{eq:energy_star_lim}) (orange dashed line). Red circles indicate the values $\gamma$=0.2, 0.3, 0.5 and 0.8.}
\label{fig:ratio}
\end{figure}

\noindent
Because $\alpha_S \propto N$, in the thermodynamic limit we can approximate
\begin{equation}
\begin{split}
G_S(0 \rightarrow \beta_m) & = \int_0^{\beta_m} dt \exp(-t)t^{\alpha_S} \\
& = \int_0^{\beta_m} dt \exp(-t + \alpha_S\ln t) \\
& \approx \int_0^{\beta_m} dt\exp(\alpha_S\ln t) = \frac{\beta_m^{\alpha_S+1}}{\alpha_S+1},
\end{split}
\end{equation}
and then we have, for the melting energy per particle $\varepsilon^* \defeq E^*/N$ in the original energy units, that
\begin{equation}
\label{eq:energy_star_lim}
\lim_{N\rightarrow \infty} \varepsilon^* = \varphi_S + \frac{\gamma}{2}+\frac{1}{2}\left(\frac{c_S+c_L}{k_B}-\gamma a_S\exp\Big(-\frac{\phi_c}{k_B T_m}\Big)\right)k_B T_m,
\end{equation}
with $\varphi_S \defeq \phi_S/N$, $c_\nu \defeq C_\nu/N$ and $a_\nu \defeq \alpha_\nu/N$. This becomes a linear relation between $E^*$ and $T_m$ for $\gamma \approx 0$, because
\begin{equation}
\lim_{\gamma \rightarrow 0} \varepsilon^* = \varphi_s + \left(\frac{c_s+c_L}{2}\right)T_m,
\end{equation}
and also for $k_B T_m \ll \phi_c$. Fig. \ref{fig:ratio} shows the dimensionless intensive quantity \[\zeta \defeq \frac{\varepsilon^*-\varphi_S}{k_B T_m}\] as a function of 
$\gamma$, using the approximation in \eqref{eq:energy_star_lim} and the exact value in \eqref{eq:energy_star}.

%
%
%

%
%
%
\section{The caloric curve in the microcanonical ensemble}
\label{sec:microcanon}

Just as we used the configurational partition function in Section ~\ref{sec:canon} to compute the canonical caloric curve, we can use the full density of states 
$\Omega(E)$ to obtain the microcanonical caloric curve, being given in terms of the CDOS as
\begin{equation}
\label{eq:etacomp}
\begin{split}
\eta(E) & = \int_0^E d\phi \mathcal{D}(\phi)\big(E-\phi\big)^{\frac{3N}{2}-1} = E^{\frac{3N}{2}-1}\int_0^E d\phi \mathcal{D}(\phi)\Big(1-\frac{\phi}{E}\Big)^{\frac{3N}{2}-1} \\
        & = E^{\frac{3N}{2}-1}\left[ \int_0^{\min(E, 1)}\hspace{-7pt} d\phi \Big(1-\frac{\phi}{E}\Big)^{\frac{3N}{2}-1}\phi^{\alpha_S}
               + \frac{\Theta(E-1)}{(1-\gamma)^{\alpha_L}}\int_1^E d\phi \Big(1-\frac{\phi}{E}\Big)^{\frac{3N}{2}-1}(\phi-\gamma)^{\alpha_L}\right] \\
        & = E^{C_S}B_S\Big(\textstyle 0 \rightarrow \min(1,\frac{1}{E})\Big) + \displaystyle\frac{\Theta(E-1)}{(1-\gamma)^{\alpha_L}}(E-\gamma)^{C_L}B_L\Big(\lambda(E) \rightarrow 1\Big)
\end{split}
\end{equation}
where we have defined the auxiliary functions
\begin{equation}
\lambda(E) \defeq \frac{1-\gamma}{E-\gamma}
\end{equation}
and
\begin{equation}
B_\nu(a \rightarrow b) \defeq \int_a^b dt\;t^{\alpha_\nu+1}(1-t)^{\frac{3N}{2}-1}
\end{equation}
for convenience of notation, where $\nu=S,L$ and $a, b \in [0, 1]$. Replacing in \eqref{eq:etacomp} we obtain $\eta(E)$ as a piecewise function,
\begin{equation}
\label{eq:eta}
\eta(E) = \begin{cases}
E^{C_S}B(\alpha_S+1, \textstyle\frac{3N}{2}) \;\text{for}\; E \leq 1\\[15pt]
\displaystyle E^{C_S}B_S(0 \rightarrow \textstyle \frac{1}{E})\displaystyle+ \frac{(E-\gamma)^{C_L}}{(1-\gamma)^{\alpha_L}}B_L(\lambda(E) \rightarrow 1)\;\text{for}\;E > 1.
\end{cases}
\end{equation}

\noindent
We can see that the microcanonical inverse temperature for the branch with $E \leq 1$ is simply given by
\begin{equation}
\beta_\text{low}(E) = \frac{\partial}{\partial E}\ln (E^{C_S}) = \frac{C_S}{E},
\end{equation}
that is $E$ as a function of $T(E)$ is a straight line with slope $C_S$, in agreement with the low temperature approximation $E_S(T)$ of the canonical caloric curve in 
(\ref{eq:solid_branch}). This means the non-monotonic behavior, i.e. the \emph{van der Waals} loop and the microcanonical melting energy $E_m$ must ocurr above $E = 1$. 
Just as we did for the canonical ensemble, we will define the probabilities of solid and liquid, $\pS{E}$ and $\pL{E}$ respectively, at an energy $E \geq 1$, according to
\begin{align}
\pS{E} & \defeq \frac{1}{\eta(E)}\int_0^1 d\phi\mathcal{D}(\phi)(E-\phi)^{\frac{3N}{2}-1} = \frac{E^{C_S}B_S(0 \rightarrow 1/E)}{\eta(E)}, \\
\pL{E} & \defeq \frac{1}{\eta(E)}\int_1^E d\phi\mathcal{D}(\phi)(E-\phi)^{\frac{3N}{2}-1} = \frac{(E-\gamma)^{C_L}}{(1-\gamma)^{\alpha_L}}\frac{B_L(\lambda(E)\rightarrow 1)}{\eta(E)}.
\end{align}

\begin{figure}
\begin{center}
\includegraphics[width=0.5\textwidth]{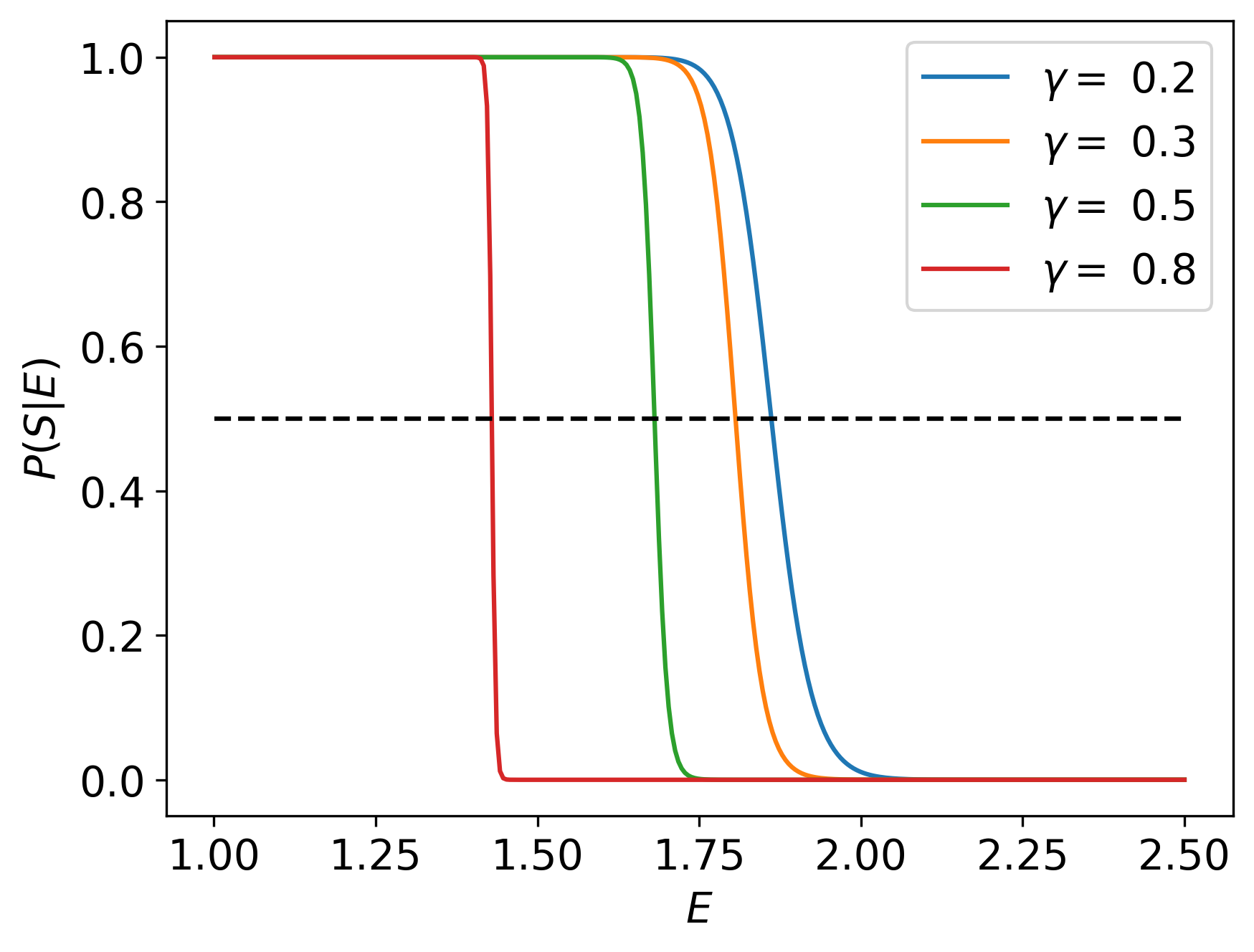}
\end{center}
\caption{Microcanonical probability of solid phase $P(S|E)$ as a function of energy, for different values of $\gamma$.}
\label{fig:probmicro}
\end{figure}

The probability of solid is shown, as a function of energy, in Fig. ~\ref{fig:probmicro}. We will define the microcanonical melting energy $E_m$ as the value of $E$ 
such that $\pS{E_m} = \pL{E_m}$, being then the solution of the transcendental equation
\begin{equation}
(E_m)^{C_S}(1-\gamma)^{\alpha_L} B_S(0 \rightarrow 1/E_m) = (E_m-\gamma)^{C_L}B_L(\lambda(E_m)\rightarrow 1).
\end{equation}

\noindent
Using the derivatives
\begin{align}
\frac{\partial}{\partial E}B_S(0 \rightarrow 1/E) & =
\frac{\partial}{\partial E}\int_0^1 dt\: t^{\alpha_S+1}(1-t)^{\frac{3N}{2}-1}\Theta(1/E-t)
= -E^{-(C_S+2)}(E-1)^{\frac{3N}{2}-1} \\
\frac{\partial}{\partial E}B_L(\lambda(E) \rightarrow 1) & =
\frac{\partial}{\partial E}\int_0^1 dt\:t^{\alpha_L+1}(1-t)^{\frac{3N}{2}-1}\Theta(t-\lambda(E))
= \frac{1}{1-\gamma}\left[\lambda(E)^{\alpha_L+3}-\lambda(E)^{C_L+2}\right],
\end{align}
we can write
\begin{equation}
\eta'(E) = \eta(E)\left(\frac{C_S}{E}\pS{E} + \frac{C_L}{E-\gamma}\pL{E}\right) + \lambda(E)^2\left[(E-\gamma)^{\frac{3N}{2}}
 -(1-\gamma)^{\frac{3N}{2}}\right] - E^{-2}(E-1)^{\frac{3N}{2}-1}
\end{equation}
and then
\begin{equation}
\beta_\text{high}(E) = \left(\frac{C_S}{E}\right)\pS{E} + \left(\frac{C_L}{E-\gamma}\right)\big(1-\pS{E}\big) + \beta_{\text{res}}(E)
\end{equation}
where
\begin{equation}
\beta_{\text{res}}(E) \defeq \frac{1}{\eta(E)}\left(\lambda(E)^2\left[(E-\gamma)^{\frac{3N}{2}}-(1-\gamma)^{\frac{3N}{2}}\right] -\frac{1}{E^2}(E-1)^{\frac{3N}{2}-1}\right)
\end{equation}
from which we see that the (inverse) temperature is also continuous at $E = 1$. Similarly to $E_{\text{res}}(\beta)$ in the canonical ensemble, we can also see that
$\beta_{\text{res}}(E)$ vanishes for both $E \rightarrow 1$ and $E \rightarrow \infty$. By using the Stirling approximation on the representation of the Beta function 
written in terms of $\Gamma$-functions,
\begin{equation}
B(x, y) = \frac{\Gamma(x)\Gamma(y)}{\Gamma(x+y)} \approx \sqrt{2\pi}\:\left[\frac{x^{x-1/2}y^{y-1/2}}{(x+y)^{x+y-1/2}}\right]
\end{equation}
for large $x$ and $y$ we have
\begin{equation}
B_\nu(a \rightarrow b) \approx
   \begin{cases}
   K_\nu \;\text{if}\; \alpha_\nu/C_\nu \in [a, b]\\[10pt]
   0\;\text{otherwise},
   \end{cases}
\end{equation}
where
\begin{equation}
K_\nu \defeq \sqrt{2\pi}\:\left[\frac{(\alpha_\nu+2)^{\alpha_\nu+3/2}{\Big(\tfrac{3N}{2}\Big)}^{3N/2-1/2}}{(C_\nu+1)^{C_\nu-1/2}}\right].
\end{equation}

%

\noindent
The probability of solid phase $\pS{E}$ is then approximated by
\begin{equation}
\pS{E} \approx \frac{E^{C_S}(1-\gamma)^{\alpha_L}K_S \mathrm{Q}[E \leq c_S/a_S]}{E^{C_S}(1-\gamma)^{\alpha_L}K_S \mathrm{Q}[E \leq c_S/a_S]
 + (E-\gamma)^{C_L} K_L \mathrm{Q}[E \geq \gamma+(c_L/a_L)(1-\gamma)]}
\end{equation}
where $\mathrm{Q}(A)$ is the indicator function~\cite{Grimmett2014} of the proposition $A$, defined as
\begin{equation}
\mathrm{Q}(A) = \begin{cases}
1\;\;\text{if}\;A\;\text{is true}, \\
0\;\;\text{otherwise}.
\end{cases}
\end{equation}

\noindent
Therefore, in this approximation the transition energy $E_m$, such that $\pS{E_m} = 1/2$, must be the solution of
\begin{equation}
E_m^{C_S}(1-\gamma)^{\alpha_L} K_S = (E_m-\gamma)^{C_L} K_L
\end{equation}
provided that no indicator function vanishes, that is, it must hold that
\begin{equation}
\gamma + \frac{c_L}{a_L}(1-\gamma) \leq E_m \leq \frac{c_S}{a_S}.
\end{equation}

\noindent
These inequalities impose a lower limit on $\gamma$, namely
\begin{equation}
\gamma \geq \frac{a_L}{a_S}-1,
\end{equation}
which however is only relevant if $a_L > a_S$, as it prevents $\gamma$ for reaching zero. In order to determine the solutions $E^*$ corresponding to the maximum 
and minimum of microcanonical temperature, we impose
\begin{equation}
\frac{\partial}{\partial E}\beta(E)\Big|_{E=E^*} = \frac{\partial^2}{\partial E^2}\ln \eta(E)\Big|_{E=E^*} = \frac{\eta''(E^*)}{\eta(E^*)}-\left(\frac{\eta'(E^*)}{\eta(E^*)}\right)^2 = 0,
\end{equation}
therefore
\begin{equation}
\label{eq:diamonds}
\eta''(E^*)\cdot \eta(E^*) = \eta'(E^*)^2.
\end{equation}

\begin{figure}
\begin{center}
\includegraphics[width=7.5cm]{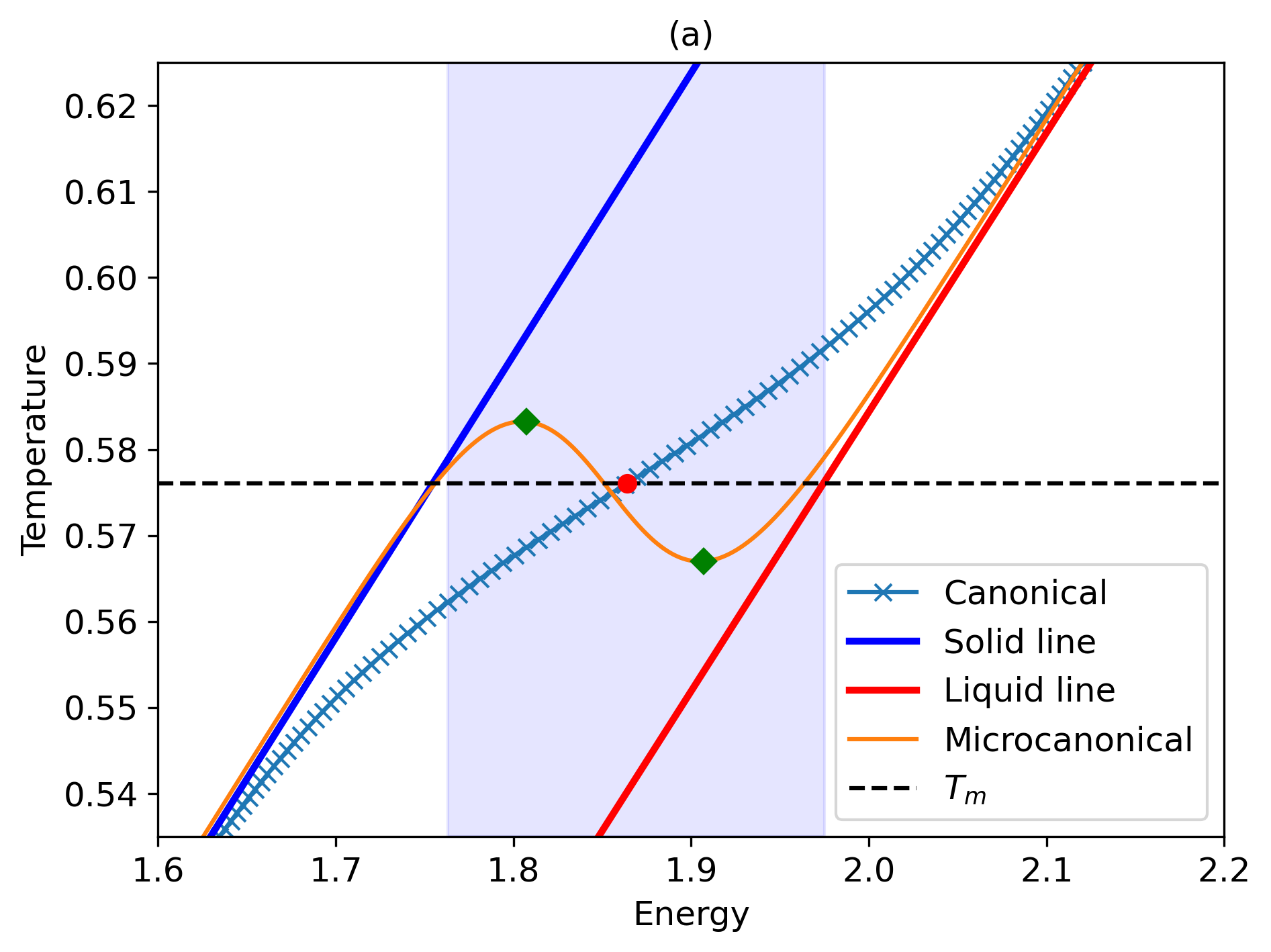}
\includegraphics[width=7.5cm]{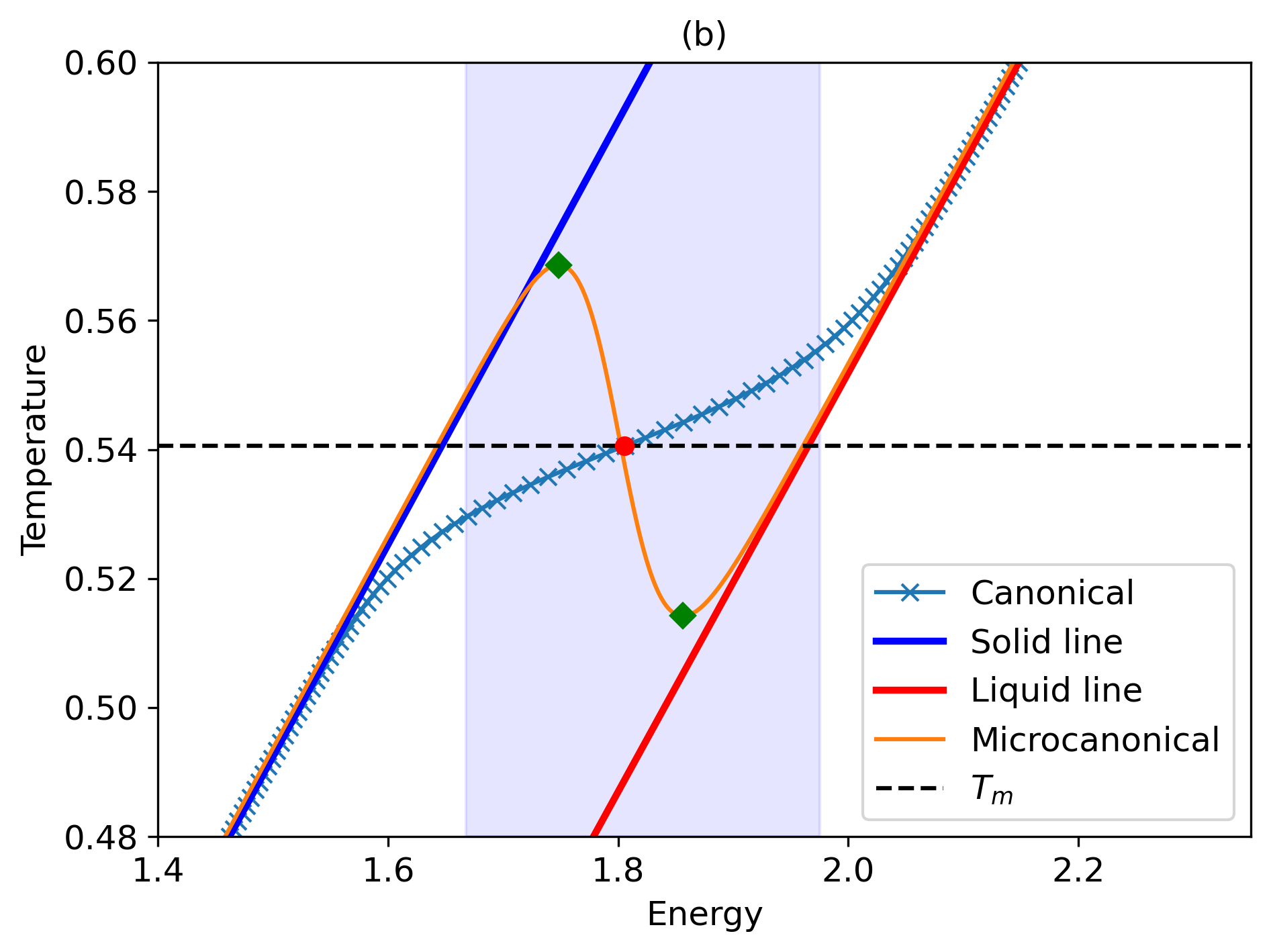}\\
\includegraphics[width=7.5cm]{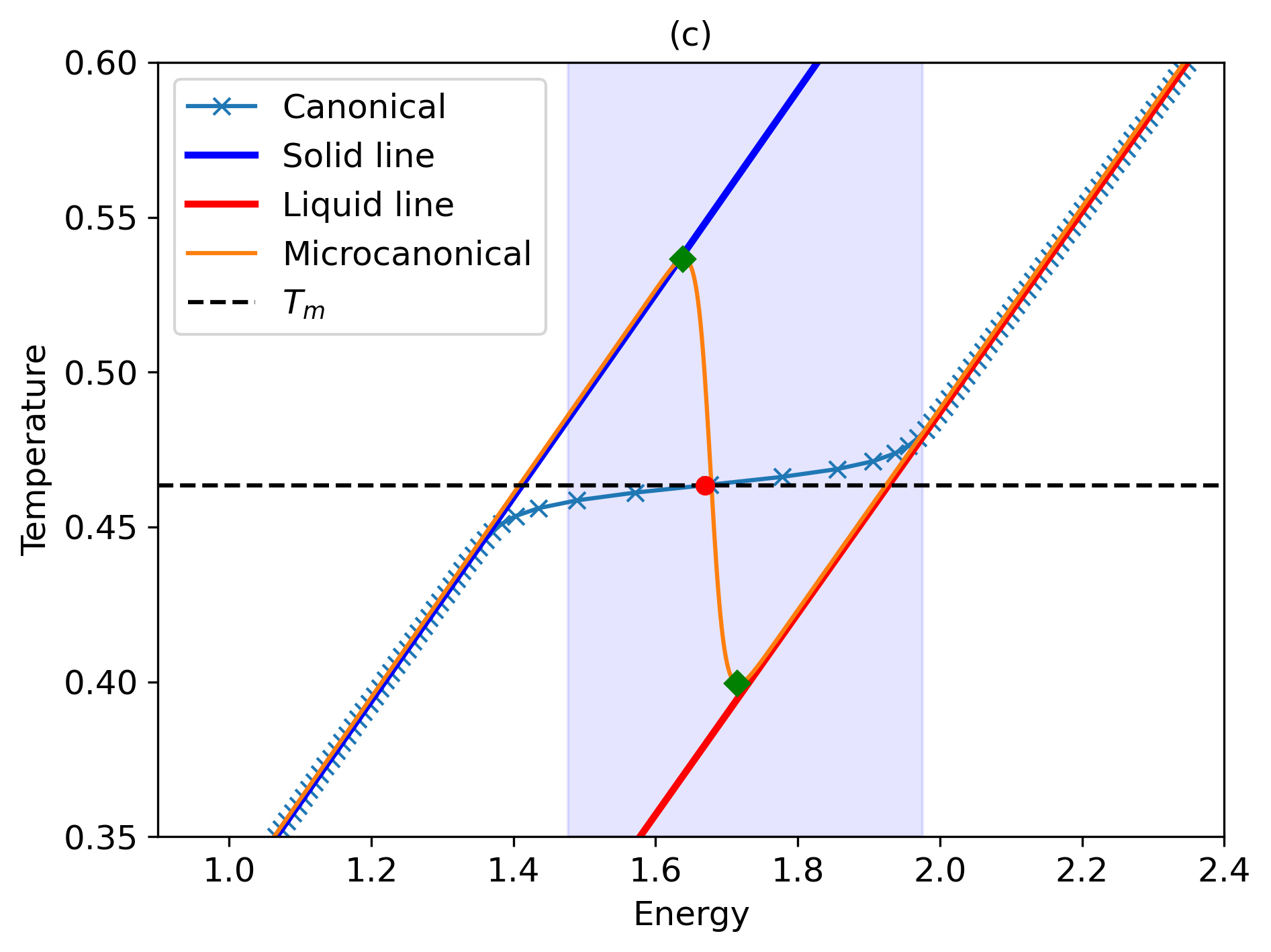}
\includegraphics[width=7.5cm]{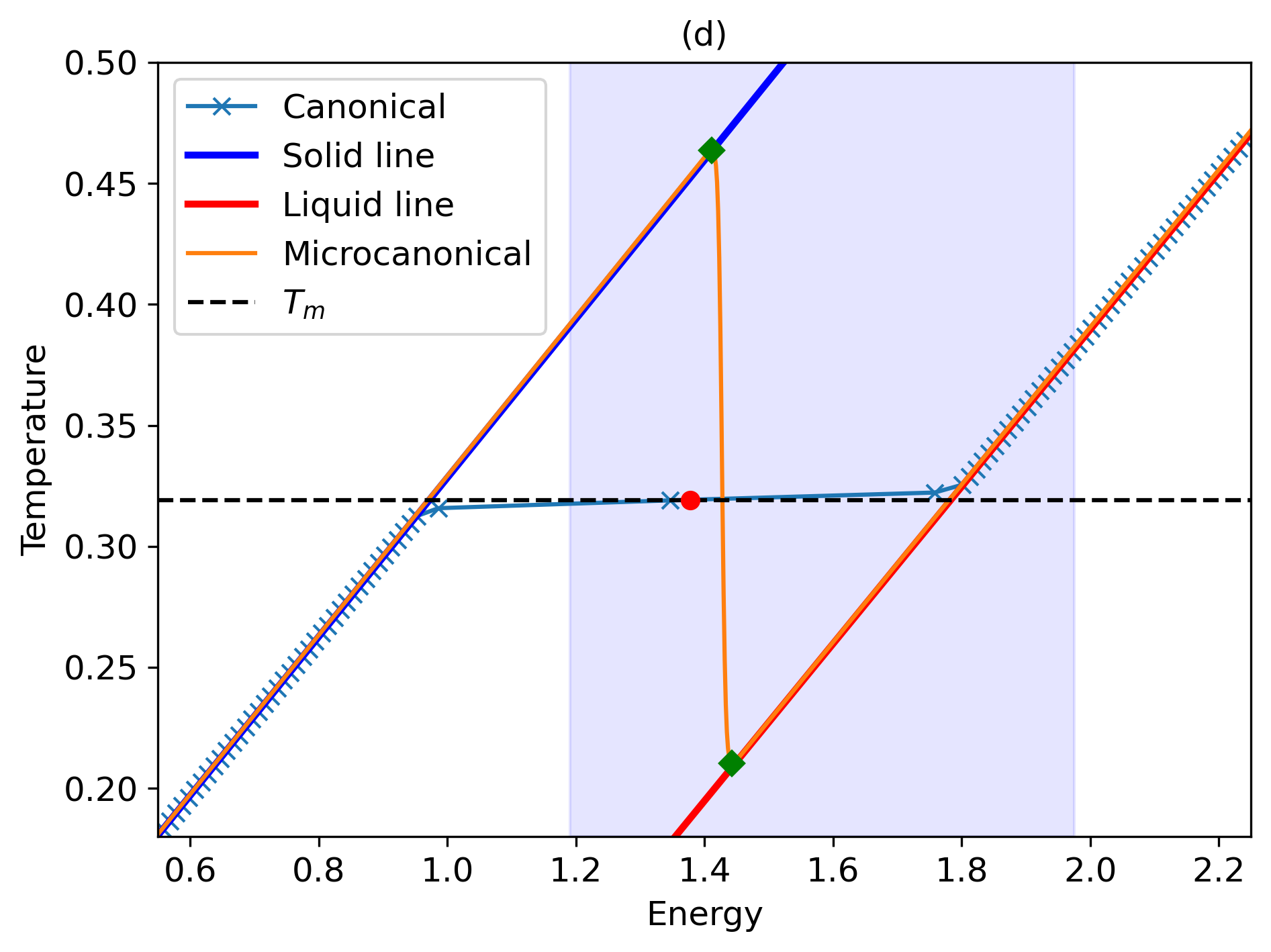}
\end{center}
\caption{Canonical and microcanonical isochoric curves for $N$=128, $\alpha_S/N$ = 1.53812 and $\alpha_L/N$ = 1.57241, with (a) $\gamma$=0.2, (b) $\gamma$=0.3, (c) $\gamma$=0.5 and (d) $\gamma$ = 0.8.}
\label{fig:isochores}
\end{figure}

The exact conditions under which \eqref{eq:diamonds} has exactly two solutions remain to be explored. Nevertheless, we have verified this fact numerically, and in Fig. 
\ref{fig:isochores} the green diamonds show the numerical solutions of (\ref{eq:diamonds}) using the expression (\ref{eq:eta}) for $\eta(E)$ in the case $E > 1$, together 
with the canonical and microcanonical caloric curves. The ensemble inequivalence expected in small systems~\cite{Dunkel2006} is clearly seen, and a rather remarkable agreement 
of the microcanonical curves with the usual shape of the Z curves in the literature is found. The degree of superheating increases with $\gamma$, and the van der Waals loop 
typically found in microcanonical curves of small systems~\cite{Doye1995, Schmidt2001, Behringer2005b, Behringer2006, Eryurek2007, Eryurek2008, Carignano2010} gradually 
becomes sharper. However, the lower inflection point in the microcanonical curve does not coincide with the value of $T_m$ obtained from the canonical curve, suggesting that the 
model for the CDOS could be improved by adding additional parameters.

%

%
%
%




%
%
%
\section{Concluding remarks}
\label{sec:conclusions}

We have presented a study of the microcanonical and canonical melting curves for a simple solid, based on the recently proposed model in Ref.~\cite{Montecinos2021}. Our 
results show that this model is sufficient to reproduce the existence of superheating and recover the so-called Z curves of microcanonical melting, in which the Z-method 
by Belonoshko \emph{et al}~\cite{Belonoshko2006} is based. This is a first step for an explanation of the foundations of the Z-method in terms of ensemble inequivalence 
for small systems.

\newpage
\bibliography{zmodel}
\bibliographystyle{unsrt}

\newpage

\end{document}